\documentclass[10pt,english,aps,prd,10pt,notitlepage,nofootinbib,preprintnumbers]{revtex4-1}
\usepackage[latin9]{inputenc}
\usepackage{amsmath}
\usepackage{amssymb}
\usepackage{esint}

\makeatletter
 
 \@ifundefined{textcolor}{}
 {%
   \definecolor{BLACK}{gray}{0}
   \definecolor{WHITE}{gray}{1}
   \definecolor{RED}{rgb}{1,0,0}
   \definecolor{GREEN}{rgb}{0,1,0}
   \definecolor{BLUE}{rgb}{0,0,1}
   \definecolor{CYAN}{cmyk}{1,0,0,0}
   \definecolor{MAGENTA}{cmyk}{0,1,0,0}
   \definecolor{YELLOW}{cmyk}{0,0,1,0}
 }

\def\Mpl{M_{\mathrm{Pl}}}

\usepackage{babel}

\usepackage{babel}

\makeatother

\usepackage{babel}
\begin{document}

\preprint{IPMU13-0178}

\title{Generalized quasi-dilaton theory}

\author{Antonio De Felice}

\email{antoniod@nu.ac.th}

\affiliation{ThEP's CRL, NEP, The Institute for Fundamental Study, Naresuan University,
Phitsanulok 65000, Thailand}

\affiliation{Thailand Center of Excellence in Physics, Ministry of Education,
Bangkok 10400, Thailand}

\author{A. Emir Gümrükçüo\u{g}lu}

\email{emir.gumrukcuoglu@ipmu.jp}

\author{Shinji Mukohyama}

\email{shinji.mukohyama@ipmu.jp}

\affiliation{Kavli Institute for the Physics and Mathematics of the Universe (WPI),
Todai Institutes for Advanced Study, University of Tokyo, 5-1-5 Kashiwanoha,
Kashiwa, Chiba 277-8583, Japan}

\begin{abstract}
Recently the first example of a unitary theory of Lorentz-invariant
massive gravity allowing for stable self-accelerating de Sitter
solutions was found, extending the quasidilaton theory. In this
paper we further generalize this new action for the quasidilaton field
by introducing general Lagrangian terms which are consistent with the 
quasidilaton symmetry while leading to second order equations of
motion. We find that the structure of the theory, compared to the
simplest stable example, does not change on introducing these new
terms.  
\end{abstract}

\maketitle

\section{Introduction}

The search for a consistent gravitational action which would lead to a
massive graviton has been pushed forward recently in several
directions. Recently, a non-linear completion of the Fierz-Pauli model
\cite{Fierz:1939ix}, which is free of the Boulware-Deser (BD) ghost
\cite{Boulware:1973my}, was introduced by de Rham, Gabadadze and Tolley
(dRGT) \cite{deRham:2010ik, deRham:2010kj} and revitalized the research
on this topic. However, shortly thereafter, the homogeneous and
isotropic solutions of this theory were found to suffer from ghost
instabilities. Specifically, the self-accelerating branch solutions
\cite{Gumrukcuoglu:2011ew} suffer from a non-linear ghost
\cite{DeFelice:2012mx}, while the remaining branch solutions have a
linear ghost instability \cite{Fasiello:2012rw} of the type found in
\cite{Higuchi:1986py}.

There has been several attempts to improve the stability of cosmological
solutions. One possibility consists of introducing inhomogeneous and/or
anisotropic background configurations either of the physical metric or
of the St\"uckelberg
fields~\cite{Chamseddine:2011bu,D'Amico:2011jj,Koyama:2011xz,Koyama:2011yg,Gratia:2012wt,Kobayashi:2012fz,Volkov:2012cf,Gumrukcuoglu:2012aa,DeFelice:2013awa}. In this approach, the deformations may stay in a hidden sector, giving a
standard Friedmann-Lema\^itre-Robertson-Walker (FLRW) form to the
physical metric, or through the recovery of the FLRW universe only in
the observable patch.

A second approach consists of introducing new degrees of freedom to the
dRGT action, in addition to the already existing gravitational
ones~\footnote{We note that all the extensions discussed here reduce to
Fierz-Pauli theory \cite{Fierz:1939ix} in the linear level. For
alternative theories which are not connected to the Fierz-Pauli
Lagrangian, see \cite{LorentzViolation}. 
}.
 One possibility is the BD ghost-free bi-metric theory, where a second
 dynamical metric is introduced and the interaction between the two
 metrics is tuned such that the BD ghost is removed by construction
 \cite{Hassan:2011zd,Hassan:2011ea,Hinterbichler:2012cn}. The properties
 of the cosmology has been studied in
 \cite{vonStrauss:2011mq,Comelli:2011zm,Akrami:2012vf,Comelli:2012db,DeFelice:2013nba}. 
As in the dRGT theory, on self-accelerating FLRW solutions three degrees
of freedom have vanishing quadratic kinetic terms and thus render those
cosmological solutions unstable at nonlinear level. On the other hand,
in the so-called normal branch without self-acceleration, stable
cosmologies are possible.

Another example to this approach is to introduce a single scalar field,
interacting with the graviton mass term. For instance, the parameters of
the dRGT theory can be promoted to vary with a dynamical scalar field
\cite{D'Amico:2011jj,Huang:2012pe}. The freedom in how these functions
vary may allow for different types of cosmologies 
\cite{Wu:2013ii,Hinterbichler:2013dv,Leon:2013qh}, and the instabilities
of the usual massive gravity can potentially be avoided
\cite{Gumrukcuoglu:2013nza}. However, the stability condition forbids
self-accelerating de Sitter solutions in this class of theories. This is
because, whenever the extra scalar field stops rolling, the system
reduces back to the original dRGT theory and thus suffers from
above-mentioned instabilities.

Recently the first example of a unitary theory of Lorentz-invariant
massive gravity with stable self-accelerating de Sitter solutions was
presented in \cite{DeFelice:2013tsa}. The theory is an extension of the
quasidilaton theory originally introduced in \cite{D'Amico:2012zv}: in
addition to the pure gravitational degrees of freedom, the Lagrangian is
endowed with an extra scalar field, which has a non-trivial coupling
with the massive graviton, and is supposed to cure some of the
unexpected pathological behavior of the original dRGT theory on
homogeneous and isotropic manifolds. The action of the system is
invariant under the so-called quasidilaton global symmetry. However, it
was shown in \cite{Gumrukcuoglu:2013nza, D'Amico:2013kya} that
self-accelerating de Sitter solutions in the original quasidilaton
theory are always plagued with a ghost degree of freedom. The theory was
thus extended in \cite{DeFelice:2013tsa} by allowing for a new coupling
(consistent with the quasidilaton symmetry) between the quasidilaton
scalar field and the St\"uckelberg fields. It is this coupling that
prevents ill-defined behavior from happening and renders the
self-accelerating solutions stable. Moreover, the inclusion of the new
coupling does not spoil the existence of the primary constraint which
removes the BD degree \cite{Mukohyama:2013raa}.

The goal of this paper is to further generalize this new action for the
quasidilaton field by introducing general Lagrangian terms which are
consistent with the quasidilaton symmetry while leading to second order
equations of motion. Some of these general terms are known in the
literature as a subset of the general Horndeski
action~\cite{Horndeski:1974wa,Deffayet:2009mn,Charmousis:2011bf}. 
We find that the structure of the theory, compared to the simplest
stable example provided in \cite{DeFelice:2013tsa}, does not change on 
introducing these new Horndeski terms. In particular, the no-ghost
condition for the degree which is cured by the inclusion of the new
coupling essentially keeps its original form upon introduction of the
Horndeski terms. The main modification appears in the expressions for
the speed of propagation and the no-ghost conditions of the remaining
degrees, which can all be satisfied within a non-null set of parameter
space.

The paper is organized as follows: In Section~II, we present the model
we consider and in Section~III we summarize the evolution equations of
the background. In Section~IV, we introduce perturbations to metric and
quasi-dilaton field and study their stability, along with several
examples. We conclude with Section~V, where we summarize our results.

\section{The model}

Let us consider the quasidilaton action, which can be written as follows \cite{D'Amico:2012zv,DeFelice:2013tsa}
\begin{equation}
S=\int d^{4}x\,\bigl[\sqrt{-g}\mathcal{L}+\Mpl^{2}m_{g}^{2}\xi e^{4\sigma/\Mpl}\,\sqrt{-\det\tilde{f}}\bigr]\,,
\end{equation}
where we have introduced the following expression 
\begin{equation}
\mathcal{L}=\mathcal{L}_{\mathrm{G}}+\mathcal{L}_{\mathrm{H}}\,.
\end{equation}
Here, $\mathcal{L}_{G}$ represents the quasidilaton dRGT Lagrangian,
that is 
\begin{equation}
\mathcal{L}_{\mathrm{G}}=\frac{\Mpl^{2}}{2}\left[R-2\Lambda+2m_{g}^{2}(\mathcal{L}_{2}+\alpha_{3}\mathcal{L}_{3}+\alpha_{4}\mathcal{L}_{4})\right],
\end{equation}
where 
\begin{eqnarray}
\mathcal{L}_{2} & \equiv & \frac{1}{2}\,([\mathcal{K}]^{2}-[\mathcal{K}^{2}])\,,\\
\mathcal{L}_{3} & \equiv & \frac{1}{6}\,([\mathcal{K}]^{3}-3[\mathcal{K}][\mathcal{K}^{2}]+2[\mathcal{K}^{3}])\,,\\
\mathcal{L}_{4} & \equiv & \frac{1}{24}\,([\mathcal{K}]^{4}-6[\mathcal{K}]^{2}[\mathcal{K}^{2}]+3[\mathcal{K}^{2}]^{2}
+8[\mathcal{K}][\mathcal{K}^{3}]-6[\mathcal{K}^{4}])\,.
\end{eqnarray}
Here, by square brackets, we indicate a trace operation, whereas $\mathcal{K}$
is the following tensor 
\begin{equation}
\mathcal{K}^{\mu}{}_{\nu}=\delta^{\mu}{}_{\nu}-e^{\sigma/\Mpl}\left(\sqrt{g^{-1}\tilde{f}}\right)_{\ \ \nu}^{\mu}\,,
\end{equation}
and 
\begin{equation}
\tilde{f}_{\mu\nu}\equiv\eta_{ab}\partial_{\mu}\phi^{a}\partial_{\nu}\phi^{b}-\frac{\alpha_{\sigma}}{\Mpl^{2}m_{g}^{2}}e^{-2\sigma/\Mpl}\partial_{\mu}\sigma\partial_{\nu}\sigma\,.\label{eq:extended-fiducial}
\end{equation}
This form of the quasidilaton dRGT Lagrangian is consistent with
the \emph{quasidilaton symmetry}, which is defined as follows \cite{D'Amico:2012zv}
\begin{equation}
\sigma\to\sigma+\sigma_{0}\,,\qquad\phi^{a}\to e^{-\sigma_{0}/\Mpl}\phi^{a}\,.
\end{equation}
Furthermore the Lagrangian is also invariant under a Poincar\'e transformation
in the space of the St\"uckelberg fields, as follows 
\begin{equation}
\phi^{a}\to\phi^{a}+c^{a}\,,\qquad\phi^{a}\to\Lambda^{a}{}_{b}\phi^{b}\,.
\end{equation}
The second term in Eq.~(\ref{eq:extended-fiducial}) gives a non-trivial
interaction term between the quasidilaton field, and the metric tensor,
which is capable, as we will see later on, to make the scalar perturbation
sector stable. In fact, we will show that the term proportional to
$\alpha_{\sigma}$ is of crucial importance to make the quasidilaton
action free of ghosts. 

Finally, we consider a general shift-symmetric Horndeski Lagrangian,
$\mathcal{L}_{\mathrm{H}}$, in the form 
\begin{equation}
\mathcal{L}_{\mathrm{H}}=P(\mathcal{X})-G_{3}(\mathcal{X})\,\Box\sigma+\mathfrak{L}_{4}+\mathfrak{L}_{5}\,,
\end{equation}
where 
\begin{eqnarray}
\mathcal{X} & \equiv & -\frac{1}{2}\, g^{\mu\nu}\,\partial_{\mu}\sigma\partial_{\nu}\sigma\,,\\
\mathfrak{L}_{4} & \equiv & G_{4,\mathcal{X}}(\Box\sigma)^{2}-G_{4,\mathcal{X}}(\nabla_{\mu}\nabla_{\nu}\sigma)\,(\nabla^{\mu}\nabla^{\nu}\sigma)+G_{4}(\mathcal{X})\, R\,,\\
\mathfrak{L}_{5} & \equiv & -\frac{G_{5,\mathcal{X}}}{6}\,(\Box\sigma)^{3}+\frac{G_{5,\mathcal{X}}}{2}\,(\nabla_{\mu}\nabla_{\nu}\sigma)\,(\nabla^{\mu}\nabla^{\nu}\sigma)\,\Box\sigma\nonumber \\
 & \qquad & {}-\frac{G_{5,\mathcal{X}}}{3}\,(\nabla^{\mu}\nabla_{\nu}\sigma)(\nabla^{\nu}\nabla_{\alpha}\sigma)(\nabla^{\alpha}\nabla_{\mu}\sigma)
{}+G_{5}(\mathcal{X})\, G_{\mu\nu}\,(\nabla^{\mu}\nabla^{\nu}\sigma)\,,
\end{eqnarray}
where the free functions $P,G_{3},G_{4},$ and $G_{5}$ are functions
of $\mathcal{X}$ only, and the subscript ``$,{\mathcal{X}}$'' denotes differentiation with respect to $\mathcal{X}$. This Lagrangian, which is invariant under
a quasidilaton symmetry transformation, has been constructed in order
to lead to, at most, second order equations of motion.

\section{The background}

On the background --- where $\sigma=\bar{\sigma}(t)$ and $\phi^a=x^a$---
the extended fiducial metric reduces to 
\begin{equation}
\tilde{f}_{00}=-\bigl(\dot{\phi}^{0}\bigr)^{2}-\frac{\alpha_{\sigma}}{\Mpl^{2}m_{g}^{2}}\, e^{-2\bar{\sigma}/\Mpl}\dot{\bar{\sigma}}^{2}\,,\qquad\tilde{f}_{ij}=\delta_{ij}\,.
\end{equation}
Then we can define the positive background variable $n$, such that
\begin{equation}
\bigl(\dot{\phi}^{0}\bigr)^{2}\equiv n(t)^{2}-\frac{\alpha_{\sigma}}{\Mpl^{2}m_{g}^{2}}\, e^{-2\bar{\sigma}/\Mpl}\dot{\bar{\sigma}}^{2}\,.\label{eq:phi0n}
\end{equation}
In other words, we have that, on the background 
\begin{equation}
\eta_{ab}\partial_{\mu}\phi^{a}\partial_{\nu}\phi^{b}=\textrm{diag}\!\left(-n^{2}+\frac{\alpha_{\sigma}}{\Mpl^{2}m_{g}^{2}}\, e^{-2\bar{\sigma}/\Mpl}\dot{\bar{\sigma}}^{2},1,1,1\right).
\end{equation}

Having introduced the variable $n$, the background for the fiducial
metric $\tilde{f}_{\mu\nu}$ is expressed in the following form 
\begin{equation}
\tilde{f}_{\mu\nu}=\textrm{diag}(-n(t)^{2},1,1,1)\,.
\end{equation}
For the background physical metric, we adopt the flat FLRW ansatz
\begin{equation}
ds^{2}=-N(t)^{2}\, dt^{2}+a(t)^{2}\delta_{ij}dx^{i}dx^{j}\,. 
 \label{eq:flatFLRW}
\end{equation}
We find it convenient to define two background variables $X$ and
$r$ as follows 
\begin{equation}
\bar{\sigma}=\Mpl\ln(a\, X)\,,\quad r\equiv\frac{n}{N}\, a\,.
\end{equation}
We consider here $a$ to be dimensionless, so as $\alpha_{\sigma}$,
$n$, $N$, $X$, $\omega$ and $r$. Also , $[\phi^{0}]=M^{-1}$,
$[H]=M$, and $[\sigma]=M$. In this case, for convenience, we can
replace the background variables $(\bar{\sigma},n)$ by means of $(X,r)$.

In the following we will consider self accelerating de Sitter solutions
for this model. It is then possible to search for solutions which
admit the following quantities to be constants: 
\begin{eqnarray}
H & \equiv & \frac{\dot{a}}{Na}\,,\\
X & \equiv & \frac{e^{\sigma/\Mpl}}{a}\,,\\
r & \equiv & \frac{n}{N}\, a\,.
\end{eqnarray}
Then we have, on the flat FLRW background (\ref{eq:flatFLRW}), that 
\begin{equation}
\mathcal{X}=\frac{1}{2}\,\frac{\dot{\sigma}^{2}}{N^{2}}=\frac{1}{2}\,\Mpl^{2}\, H^{2}=\textrm{constant}\,.
\end{equation}
The Friedmann equation reads 
\begin{eqnarray}
\Lambda & = & 3{H}^{2}+\frac{P}{\Mpl^{2}}-{H}^{2}P_{{,\mathcal{X}}}-3G_{{3,\mathcal{X}}}\,{H}^{4}{\Mpl}\nonumber \\
 & + & [3(X-1)(X-2)-(X-1)^{2}(X-4)\alpha_{{3}}-(X-1)^{3}\alpha_{{4}}]m_{g}^{2}\nonumber \\
 & - & 12G_{{4,\mathcal{X}}}\,{H}^{4}-6{H}^{6}{M}^{2}G_{{4,\mathcal{XX}}}+6\,G_{4}\,\frac{{H}^{2}}{\Mpl^{2}}\nonumber \\
 & - & {H}^{8}{M}^{3}G_{{5,\mathcal{XX}}}-5\,G_{{5,\mathcal{X}}}\,{H}^{6}{\Mpl}\,.\label{eq:fried}
\end{eqnarray}
Looking for the condition of a positive background effective gravitational
constant we can impose the relation 
\begin{equation}
\frac{\partial\Lambda}{\partial(H^{2})}>0\,,
\end{equation}
which leads to the condition 
\begin{eqnarray}
P_{,\mathcal{X}} & < & 6-6{H}^{2}\left(5G_{{5,\mathcal{X}}}\,{H}^{2}+2G_{{3,\mathcal{X}}}\right)\Mpl+{\frac{12G_{4}}{{\Mpl^{2}}}}
-{H}^{2}\left(48\,{H}^{2}G_{{4,\mathcal{XX}}}+P_{,\mathcal{XX}}\right){\Mpl^{2}}-42{H}^{2}G_{{4,\mathcal{X}}}\nonumber \\
 && -  {\Mpl^{5}}{H}^{8}G_{{5,\mathcal{XXX}}}-6{\Mpl^{4}}{H}^{6}G_{{4,\mathcal{XXX}}}
-{H}^{4}\left(13\,{H}^{2}G_{{5,\mathcal{XX}}}+3G_{{3,\mathcal{XX}}}\right){\Mpl^{3}}\,.
\end{eqnarray}
In the case $P_{,\mathcal{X}}=\omega$, and in the absence of the
other Horndeski terms one finds the condition $\omega<6$. If, for
example, we add a rather simple Horndeski term, namely the cubic
galileon term $G_{3}=-\tilde{g}_{3}\mathcal{X}/(\Mpl m_{g}^{2})$,
we find $\omega<6+12\tilde{g}_{3}H^{2}/m_{g}^{2}$.

Besides the Friedmann equation, there are other two independent equations.
We can choose them, for example, to be the second Einstein equation
and the equation of motion for the scalar field $\sigma$. The variation
of the action with respect to the St\"uckelberg fields does
 not introduce
new independent equations of motion. On solving the above-mentioned
three independent equations of motion, we can constrain three parameters.
One constraint is given by the Friedmann equation (\ref{eq:fried}).
Another one can be written as 
\begin{equation}
3(X-1)-3(X-1)^{2}\alpha_{3}+(X-1)^{3}\alpha_{4}+\xi X^{3}=0\,.\label{eq:constr2}
\end{equation}
Notice that this is the term which multiplies the quantity $\alpha_{\sigma}$
in the equation of motion for the scalar field $\sigma$. Therefore
$\alpha_{\sigma}$ never enters the equations of motion for the self
accelerating backgrounds of this model. Finally, we can write the
$\sigma$ equation of motion which gives the last independent constraint
as 
\begin{eqnarray}
(X-1)(r-1)X^{2}m_{g}^{2}\alpha_{3} & = & 3{H}^{4}\Mpl G_{{3,\mathcal{X}}}+2\left(r-1\right){X}^{2}m_{g}^{2}
 + 6{H}^{6}{\Mpl^{2}}G_{{4,\mathcal{XX}}}+3{H}^{6}\Mpl G_{{5,\mathcal{X}}}\nonumber \\
 && +  {H}^{8}{\Mpl^{3}}G_{{5,\mathcal{XX}}}+6{H}^{4}G_{{4,\mathcal{X}}}
  +  {H}^{2}P_{,\mathcal{X}}+\frac{{X}^{4}(r-1)\xi m_{g}^{2}}{X-1}\,.\label{eq:constr3}
\end{eqnarray}

In the following, we shall use equations (\ref{eq:fried}), (\ref{eq:constr2}),
and (\ref{eq:constr3}) in order to replace the constants $\Lambda$,
$\alpha_{3}$, and $\alpha_{4}$ in terms of the other constants/parameters
of the model.

\section{Perturbations}

In order to make sure that the de Sitter solutions are stable and
do not lead to pathological degrees of freedom, we need to study the
behavior of the perturbations fields around such backgrounds.

\subsection{Scalar perturbations}

We work here in the unitary gauge, where the St\"uckelberg fields are
not perturbed. This choice completely fixes the gauge for the scalar,
vector and tensor modes.

As for the scalar sector we introduce the metric in the form 
\begin{equation}
\delta g_{00}  =  -2\,N^{2}\Phi\,,\qquad
\delta g_{0i} =  N\,a\,\partial_{i}B\,,\qquad
\delta g_{ij}  =  a^{2}\left[2\delta_{ij}\Psi+\left(\partial_{i}\partial_{j}-\frac{1}{3}\,\delta_{ij}\partial_{l}\partial^{l}\right)E\right]\,,
\end{equation}
whereas the dilaton field is perturbed as 
\begin{equation}
\sigma=\bar{\sigma}+\Mpl\,\delta\sigma\,.
\end{equation}

In order to simplify the analysis we introduce the following
constants 
\begin{eqnarray}
&&G_{3}(\mathcal{X})  \equiv  g_{3}\Mpl\,,\qquad\;\,
G_{3,\mathcal{X}}  \equiv \frac{g_{3x}}{\Mpl H^{2}}\,,\quad\quad
G_{3,\mathcal{XX}}  \equiv \frac{g_{3xx}}{\Mpl^{3}H^{4}}\,,
\nonumber\\
&&G_{4}(\mathcal{X})  \equiv  \Mpl^{2}g_{4}\,,\qquad\;\,
G_{4,\mathcal{X}}  \equiv  \frac{g_{4x}}{H^{2}}\,,\qquad\quad\;\;\,
G_{4,\mathcal{XX}}  \equiv  \frac{g_{4xx}}{\Mpl^{2}H^{4}}\,,\qquad
G_{4,\mathcal{XXX}}  \equiv  \frac{g_{4xxx}}{\Mpl^{4}H^{6}}\,,
\nonumber\\
&&G_{5}(\mathcal{X})  \equiv \frac{\Mpl}{H^{2}}\, g_{5}\,,\qquad
G_{5,\mathcal{X}} \equiv \frac{g_{5x}}{\Mpl H^{4}}\,,\qquad
G_{5,\mathcal{XX}}  \equiv  \frac{g_{5xx}}{\Mpl^{3}H^{6}}\,,\qquad
G_{5,\mathcal{XXX}}  \equiv  \frac{g_{5xxx}}{\Mpl^{5}H^{8}}\,,
\nonumber\\
&&P(\mathcal{X})  \equiv  p\,\Mpl^{2}H^{2}\,,\qquad
P_{,\mathcal{X}}  \equiv  p_{x}\,,\qquad
P_{,\mathcal{XX}}  \equiv  \frac{p_{xx}}{\Mpl^{2}H^{2}}\,,\qquad
\alpha_{\sigma} \equiv  \frac{m_{g}^{2}X^{2}}{H^{2}}\bar{\alpha}\,,\qquad
m_{g}^{2}\xi  \equiv  \bar{\xi}H^{2}\,.
\end{eqnarray}
On using these variables, together with Eq.~(\ref{eq:phi0n}), we
find that, on the self-accelerating backgrounds, 
\begin{equation}
\left(\frac{\dot{\phi}^{0}}{n}\right)^{2}=1-\frac{\bar{\alpha}}{r^{2}}>0\,,
\end{equation}
which implies 
\begin{equation}
\bar{\alpha}<r^{2}\,.\label{eq:alpr}
\end{equation}
This condition defines a set of consistent background variables. Although
this condition does not constrain any parameter space in the simplest
case (as we will see later on), in general, it will restrict the allowed
parameter space for more general models.

\subsubsection{No-ghost conditions}

On expanding the action at second order in the perturbation fields,
we can integrate out the fields $B$ and $\Phi$, as usual. Furthermore,
because of the structure of the gravitational Lagrangian, we find that,
on introducing the field redefinition 
\begin{equation}
\delta\sigma=\Psi+\bar{\delta\sigma}\,,
\end{equation}
the field $\Psi$ also becomes an auxiliary field. After integrating
out the field $\Psi$ (the would be Boulware-Deser ghost), the theory
only admits two propagating scalar fields. By studying the property
of the kinetic matrix in the total Lagrangian $\mathcal{L}\ni K_{11}|\dot{\bar{\delta\sigma}}|^{2}+K_{22}|\dot{E}|^{2}+K_{12}(\dot{\bar{\delta\sigma}}^{\dagger}\dot{E}+\mathrm{h.c.})$,
we find that, in order to remove any ghost degree of freedom, for
any $k$-mode, we require the following two conditions to hold 
\begin{eqnarray}
K_{22} & = & \frac{a^{4}\gamma_{1}Hk^{4}\Mpl^{2}}{36\dot{a}}\left[(\bar{\alpha}-1)\gamma_{4}\frac{k^{2}}{a^{2}H^{2}}+3\gamma_{2}\gamma_{3}\right] \left[(\bar{\alpha}-1)\gamma_{5}^{2}\frac{k^{2}}{a^{2}H^{2}}+\gamma_{2}\gamma_{3}\right]^{-1}>0\,,\nonumber\\
\det(K_{IJ}) & = & \left[\frac{\bar{\alpha}\gamma_{4}(r-1)^{2}\frac{k^{2}}{a^{2}H^{2}}+3\gamma_{2}\gamma_{3}(r^{2}-\bar{\alpha})}{(\bar{\alpha}-1)\gamma_{5}^{2}\frac{k^{2}}{a^{2}H^{2}}+\gamma_{2}\gamma_{3}}\right]
\left[ \frac{a^{10}\gamma_{1}\gamma_{2}H^{4}k^{2}\Mpl^{4}}{8\dot{a}^{2}(r-1)^{2}r^{2}}\right]>0\,,
\end{eqnarray}
where 
\begin{eqnarray}
\gamma_{1} & \equiv & 1-2g_{4x}-g_{5x}+2g_{4}\,,\nonumber\\
\gamma_{2} &\equiv& 3g_{3x}+6g_{4x}+3g_{5x}+6g_{4xx}+g_{5xx}+p_{x}\,,\nonumber\\
\gamma_{2x} &\equiv& 3g_{3xx}+6g_{4xx}+3g_{5xx}+6g_{4xxx}+g_{5xxx}+p_{xx}\,,\nonumber\\
\gamma_3&\equiv& 6+12\,g_4-9\,(2\,\gamma_1+\gamma_5)-(\gamma_2+\gamma_{2x})\,,\nonumber\\
\gamma_{4} & \equiv & 3\gamma_{5}^{2}-2\gamma_{1}\gamma_{3}\,,\nonumber\\
\gamma_{5} &\equiv& g_{3x}+8g_{4x}+5g_{5x}+4g_{4xx}+g_{5xx}-4g_{4}-2\,,
\end{eqnarray}
We notice that 
\begin{equation}
\frac{\partial\Lambda}{\partial(H^{2})}=\frac{\gamma_{3}}{2}>0\,,
\end{equation}
so that, we need to impose the following conditions 
\begin{equation}
K_{22}>0\,,\quad\det(K_{IJ})>0\,,\quad\gamma_{3}>0\,,
\end{equation}
and, by requiring the result to be independent of the value of the
wave vector $k$, we need to further impose for $K_{22}$ 
\begin{equation}
\frac{(\bar{\alpha}-1)\gamma_{4}}{\gamma_{2}}>0\,,\quad\frac{(\bar{\alpha}-1)}{\gamma_{2}}>0\,,\quad\gamma_{1}>0\,,
\end{equation}
which together impose $\gamma_{4}>0$. Using (\ref{eq:alpr}), the
positivity of the determinant yields 
\begin{equation}
\frac{\bar{\alpha}\gamma_{4}}{\gamma_{2}}>0\,.
\end{equation}
Collecting all the conditions, the allowed parameter region is 
\begin{equation}
\gamma_{1}>0\,,\quad\gamma_{2}>0\,,\quad\gamma_{3}>0\,,\quad\gamma_{4}>0\,,\quad r>1\,,\quad1<\bar{\alpha}<r^{2}\,.\label{eq:no-ghost-scalar}
\end{equation}

Notice though, that for the general model we need a positive (non-zero,
in particular) value for $\bar{\alpha}$ (i.e.\ $\alpha_{\sigma}$)
in order not to have ghosts in the scalar sector. This no-ghost conditions,
does not depend on any of the new Horndeski terms, so that this same
condition applies, unchanged, to the simplest~\cite{DeFelice:2013tsa},
as well as to the most complicated theory of these models.

\subsubsection{Speed of propagation. }

In order to find the speed of propagation for the scalar modes, we
find it convenient to diagonalize the kinetic matrix $K_{IJ}$ by
defining the fields $q_1$ and $q_2$ as
\begin{equation}
\bar{\delta s}  \equiv  k\, q_{1}\,,\qquad
E  \equiv  \frac{q_{2}}{k^{2}}-\frac{K_{12}}{K_{22}}\, k\, q_{1}\,.
\end{equation}
The $k$-dependence in this field redefinition has been introduced
so that, for the new kinetic matrix, the diagonal elements tend to
finite (non-zero) values for large $k$'s.

The new kinetic matrix $\mathcal{T}_{IJ}$ can be written, without
approximations, as 
\begin{equation}
\mathcal{L}\ni\mathcal{T}_{11}(t,k)\,|\dot{q}_{1}|^{2}+\mathcal{T}_{22}(t,k)|\dot{q}_{2}|^{2}\,,
\end{equation}
and, when the no-ghost conditions hold we consistently find 
\begin{equation}
\mathcal{T}_{11}>0\,,\qquad\textrm{and}\qquad\mathcal{T}_{22}>0\,.
\end{equation}

For large momenta (with respect to $H$ and $m_{g}$), the structure
of the equations of motion for the total Lagrangian can be approximated
as 
\begin{eqnarray}
\mathcal{T}_{11}\,\ddot{q}_{1}+k\,\mathcal{B}\,\dot{q}_{2} =0\,,\qquad
\mathcal{T}_{22}\,\ddot{q}_{2}-k\,\mathcal{B}\,\dot{q}_{1}+k^{2}\mathcal{C}q_{2} = 0\,,
\end{eqnarray}
with 
\begin{eqnarray}
\mathcal{T}_{11} & \approx & \frac{9}{2}\,\frac{\gamma_{2}\bar{\alpha}H^{3}\Mpl^{2}a^{6}}{r^{2}\dot{a}(\bar{\alpha}-1)}\,,\nonumber\\
\mathcal{T}_{22} & \approx & \frac{a^{4}\gamma_{1}\gamma_{4}H\Mpl^{2}}{36\dot{a}\gamma_{5}^{2}}\,,\nonumber\\
\mathcal{B} & \approx & \frac{a^{3}H\Mpl^{2}\bar{\alpha}\gamma_{1}\gamma_{2}}{2r(1-\bar{\alpha})\gamma_{5}}\,,\nonumber\\
\mathcal{C} & \approx & 
-\frac{\dot{a}\Mpl^{2}}{36(\bar{\alpha}-1)\gamma_{5}^{2}H}\,\Big\{2\gamma_{1}^{2}[(\bar{\alpha}-1)\gamma_{5}+\gamma_{2}]+(\bar{\alpha}-1)\gamma_{5}^{2}\gamma_{6}\Big\}\,,
\end{eqnarray}
where now $\mathcal{T}_{11}$, $\mathcal{T}_{22}$, $\mathbf{\mathcal{B}}$,
and $\mathcal{C}$ are $k$-independent, as only their leading order
term in large $k$-expansion has been considered here. All other terms
in the equations of motion are suppressed by inverse powers of $k/(aH)$
and/or $k/(am_{g})$. We note that this approximation breaks down when
$\bar{\alpha}-1={\cal O}(a\, H/k)$. Otherwise, the large $k$-expansion
employed here is justified deep inside the horizon.

Then, the speed of propagation of one of the two scalar modes reduces to 
\begin{equation}
c_{s}^{2}=\frac{\mathcal{B}^{2}+\mathcal{C}\,\mathcal{T}_{11}}{\mathcal{T}_{11}\mathcal{T}_{22}}\,\frac{a^{2}}{N^{2}}
=\frac{2\gamma_{1}^{2}(\gamma_{2}-\gamma_{5})-\gamma_{5}^{2}\gamma_{6}}{\gamma_{1}\gamma_{4}}\,,
\end{equation}
where 
\begin{equation}
\gamma_{6}\equiv2g_{4}+1\,.
\end{equation}
The other scalar mode has vanishing sound speed, and this property is
not affected by the introduction of the new Horndeski terms. Hence, the
scalar sector does not have any instabilities whose time-scales are
parametrically shorter than the background cosmological time-scale if
the condition (\ref{eq:no-ghost-scalar}) and 
\begin{equation}
 c_s^2 > 0
\end{equation} 
are satisfied.

\subsection{Tensor perturbations}

The action for the tensor perturbation modes reduces to 
\begin{equation}
\mathcal{L}_{GW}=\frac{\Mpl^{2}}{8}a^{3}N\gamma_{1}\left[\frac{|\dot{h}_{ij}|^{2}}{N^{2}}-\left(c_{G}^{2}\frac{k^{2}}{a^{2}}+M_{GW}^{2}\right)|h_{ij}|^{2}\right].
\end{equation}

The no-ghost condition for the tensor modes is then 
\begin{equation}
\gamma_{1}>0\,.
\end{equation}
The speed of propagation, for large $k$, for the tensor modes becomes
\begin{equation}
c_{G}^{2}=\frac{\gamma_{6}}{\gamma_{1}}\,,
\end{equation}
and its mass is 
\begin{equation}
M_{GW}^{2}={\frac{(r-1){X}^{3}m_{g}^{2}}{\gamma_{1}\left(X-1\right)}}+{\frac{(r-1){X}^{4}{H}^{2}\bar{\xi}}{\gamma_{1}\left(X-1\right)^{2}}}+{\frac{(Xr+r-2)\gamma_{2}{H}^{2}}{\gamma_{1}\left(r-1\right)\left(X-1\right)}}\,.
\end{equation}
Since $M_{GW}$ is generically of order $|m_g|\simeq H$,
 the tensor sector does not have any instabilities whose time-scales are parametrically
shorter than the background cosmological time-scale if 
\begin{equation}
c_{G}^{2}>0. 
\end{equation}

\subsection{Vector perturbations}

The reduced action for the vector modes reads as 
\begin{equation}
\mathcal{L}_{V}=\frac{\Mpl^{2}}{16}a^{3}N\gamma_{1}\left[\frac{Q_{V}|\dot{E}_{i}|^{2}}{N^{2}}-k^{2}M_{GW}^{2}|E_{i}|^{2}\right].
\end{equation}
As for the vector modes, after imposing $\gamma_{1}>0$, we have the
no-ghost condition 
\begin{equation}
Q_{V}\equiv\frac{2\gamma_{2}k^{2}}{\gamma_{1}(r^{2}-1)\frac{k^{2}}{H^{2}a^{2}}+2\gamma_{2}}>0\,,
\end{equation}
which is always satisfied, for the no-ghost parameter space allowed
by the scalar modes. The speed of propagation, for large $k$, is
\begin{equation}
c_{V}^{2}=\frac{\gamma_{1}M_{GW}^{2}(r^{2}-1)}{2H^{2}\gamma_{2}}\,.
\end{equation}
This expression has an interesting consequence: since $r>1$ and
$\gamma_{1,2}>0$, the absence of gradient instability in the vector
sector fixes the squared-mass of the tensor modes to be positive, i.e.
\begin{equation}
M_{GW}^{2}>0. 
\end{equation}

\subsection{Allowed parameter space}

The above conditions all together give the following allowed parameter
space 
\begin{eqnarray}
\gamma_{1} & > & 0\land M_{GW}^{2}>0\land0<\gamma_{3}<3\gamma_{5}^{2}/(2\gamma_{1})\nonumber \\
 & \land & 0<\gamma_{6}<2\gamma_{1}^{2}(\gamma_{2}-\gamma_{5})/\gamma_{5}^{2}\nonumber \\
 & \land & 1<\bar{\alpha}<r^{2}\land r>1\land\gamma_{2}>0\,.\label{eq:condo}
\end{eqnarray}
This is tantamount to saying that the set of parameter space for the
models which have a well-behaved stable late time de Sitter solution,
is not empty.

\subsection{Examples}

\subsubsection{K-essence like case}

The first example consists of setting to zero any Horndeski term,
as well as the non-derivative coupling ($\xi=0$). In this case we
find 
\begin{equation}
\gamma_{1}=1\,,\qquad\gamma_{2}=p_{x}\,,\qquad\gamma_{3}=6-p_{x}-p_{xx}\,,\qquad\gamma_{5}=-2\,,\qquad\gamma_{6}=1\,,
\end{equation}
so that the allowed parameter space (\ref{eq:condo}) becomes
\begin{equation}
p_{x}>0\land0<p_{x}+p_{xx}<6\land M_{GW}^{2}>0\land r>1\land1<\bar{\alpha}<r^{2}\,,
\end{equation}
or, explicitly 
\begin{eqnarray}
r & > & 1\land1<\bar{\alpha}<r^{2}\land p_{x}>0\land0<p_{x}+p_{xx}<6\nonumber \\
 & \land & \left[\left(0<X<1\land m_{g}^{2}<-\frac{H^{2}p_{x}(r+rX-2)}{X^{3}(r-1)^{2}}\right)\right.\nonumber \\
 & \lor & \left.\left(X>1\land m_{g}^{2}>-\frac{H^{2}p_{x}(r+rX-2)}{X^{3}(r-1)^{2}}\right)\right].
\end{eqnarray}
The speed of propagation of one scalar mode is 
\begin{equation}
c_{s}^{2}=\frac{p_{x}}{p_{x}+p_{xx}}\,,
\end{equation}
whereas the other scalar mode has vanishing speed of propagation.
The speed of propagation of the tensor modes is unity, whereas the
one of the vector modes reduces to 
\begin{equation}
c_{V}^{2}=\frac{M_{GW}^{2}(r^{2}-1)}{2H^{2}p_{x}}\,,
\end{equation}
where 
\begin{equation}
M_{GW}^{2}=\frac{(r-1){X}^{3}m_{g}^{2}}{X-1}+\frac{(Xr+r-2)p_{x}{H}^{2}}{\left(r-1\right)\left(X-1\right)}\,.
\end{equation}
For the simplest case, $p_{xx}=0$, and $p_{x}=\omega$, we confirm
the results as given in \cite{DeFelice:2013tsa}.

\subsubsection{Horndeski case}

Let us consider one of the easiest generalization from the Horndeski
Lagrangian. Setting $g_{4}$, $g_{5}$, their derivatives and $\xi$
to zero, let us assume $p_{x}=\omega$, $p_{xx}=0$, and $g_{3xx}=0$.
Then, we have 
\begin{equation}
\gamma_{1}=1\,,\qquad\gamma_{2}=\omega+3g_{3x}\,,\qquad\gamma_{3}=6-\omega-12g_{3x}\,,\qquad\gamma_{5}=g_{3x}-2\,,\qquad\gamma_{6}=1\,,
\end{equation}
so that, we find the following the set of allowed parameter space,
namely, 
\begin{eqnarray}
r & > & 1\land12g_{3x}+\omega<6\land1<\bar{\alpha}<r^{2}\nonumber \\
 & \land & \left[\left(0<X<1\land m_{g}^{2}<-\frac{H^{2}(\omega+3g_{3x})(r+rX-2)}{X^{3}(r-1)^{2}}\right)\right.\nonumber \\
 & \lor & \left.\left(X>1\land m_{g}^{2}>-\frac{H^{2}(\omega+3g_{3x})(r+rX-2)}{X^{3}(r-1)^{2}}\right)\right]\nonumber \\
 & \land & [(\omega+3g_{3x}>0\land0<g_{3x}<2/3)\nonumber \\
 & \lor & (-8-2\sqrt{19}<g_{3x}\leq-1\land g_{3x}(g_{3x}-8)<2\omega)\nonumber \\
 & \lor & (-1<g_{3x}\leq0\land2\omega+3g_{3x}(4+g_{3x})>0)]\,,
\end{eqnarray}
which implies $-2(4+\sqrt{19})<g_{3x}<2/3$.

In this case, the speed of the propagating scalar mode is 
\begin{equation}
c_{s}^{2}=1-\frac{4g_{3x}(1+g_{3x})}{2\omega+3g_{3x}(4+g_{3x})}\,,
\end{equation}
which is superluminal if $-1<g_{3x}<0$. The vector modes propagate
with speed 
\begin{equation}
c_{V}^{2}=\frac{M_{GW}^{2}(r^{2}-1)}{2H^{2}(\omega+3g_{3x})}\,,
\end{equation}
where 
\begin{equation}
M_{GW}^{2}={\frac{(r-1){X}^{3}m_{g}^{2}}{X-1}}+{\frac{(Xr+r-2)(\omega+3g_{3x}){H}^{2}}{\left(r-1\right)\left(X-1\right)}}\,.
\end{equation}
The tensor modes, on the other hand, propagate with unity speed of
propagation.

\subsubsection{Vanishing bare cosmological constant}

Let us consider the case of a vanishing bare cosmological constant.
In this case, if a de Sitter solution exists, the system will be self-accelerating.
The condition $\Lambda=0$, sets a constraint between $m_{g}^{2}$
and the other variables, as in 
\begin{equation}
\frac{m_{g}^{2}}{H^{2}}=\frac{\frac{\gamma_{2}[X(rX-2)+1]}{(r-1)X^{2}}-(2\gamma_{1}+\gamma_{6}+p+\bar{\xi}X^{2}-2g_{4x})}{(X-1)^{2}}\,.
\end{equation}
On inserting such relation in the expression for $m_{g}^{2}$ into
the expression of $M_{GW}^{2}$, we find 
\begin{equation}
M_{GW}^{2}  = \frac{H^{2}}{\gamma_{1}(r-1)(X-1)^{3}}\,\Big[\gamma_{2}(r^{2}X^{3}-3rX^{2}+r+3X-2) - (2\gamma_{1}+\gamma_{6}+\gamma_{7})(r-1)^{2}X^{3}\Big]\,,
\end{equation}
where 
\begin{equation}
\gamma_{7}=p+\bar{\xi}X-2g_{4x}\,.
\end{equation}
Then, on defining 
\begin{eqnarray}
\Gamma \equiv (2\gamma_{1}+\gamma_{6}+\gamma_{7})(r-1)^{2}X^{3} - \gamma_{2}\{r[X^{2}(rX-3)+1]+3X-2\}\,,
\end{eqnarray}
we find the following allowed parameter space 
\begin{eqnarray}
0 & < & \gamma_{3}<\frac{3\gamma_{5}^{2}}{2\gamma_{1}}\land0<\gamma_{6}<\frac{2\gamma_{1}^{2}\left(\gamma_{2}-\gamma_{5}\right)}{\gamma_{5}^{2}}\nonumber \\
 & \land & r>1\land1<\bar{\alpha}<r^{2}\land\{\gamma_{1}>0\land\gamma_{2}>0\nonumber \\
 & \land & [(\Gamma<0\land X>1)\lor(\Gamma>0\land0<X<1)]\}\,.
\end{eqnarray}

\section{Discussion and Conclusions}

We have studied a form for the quasidilaton action, which generalizes
the recent ghost-free quasidilaton action introduced in \cite{DeFelice:2013tsa}. We
have found that all these model possess, in general, self-accelerating
solutions. The background dynamics of the de Sitter solutions does
\emph{not} depend on $\alpha_{\sigma}$. Therefore the background
evolution is exactly the same as the original quasidilaton case already
introduced in \cite{D'Amico:2012zv}. Nonetheless, this same parameter
heavily affects the stability of the perturbation fields.

We have found that the background condition $(\dot{\phi}^{0}/n)^{2}>0$, 
implies 
\begin{equation}
\frac{\alpha_{\sigma}H^{2}}{m_{g}^{2}}<r^{2}X^{2}\,.
\end{equation}
This condition is restrictive enough to make the general model have
the same structure of the simplest allowed case introduced in \cite{DeFelice:2013tsa}.
Furthermore, the no-ghost conditions for the scalar sector impose,
in general, 
\begin{equation}
r>1\,,\quad X^{2}<\frac{\alpha_{\sigma}H^{2}}{m_{g}^{2}}<r^{2}X^{2}\,,
\end{equation}
so that the parameter $\alpha_{\sigma}/m_g^2$
 needs to be positive (different from zero, in particular).

A coupling similar to $\alpha_{\sigma}$ exists also in the DBI Galileon
coupled to massive gravity (DBI massive
gravity)~\cite{Gabadadze:2012tr}. Among various Lagrangian terms of the
generalized quasidilaton theory considered in the present paper, some
are allowed in the DBI massive gravity as well but some are
forbidden. Specifically, the parameter space (\ref{eq:condo}) does not
include a region corresponding to the DBI massive gravity.

In the allowed parameter space defined in (\ref{eq:condo}), all the
expected perturbation modes are well behaved: they possess positive
kinetic energy and non-negative squared speed of propagation. One
scalar mode has always zero speed of propagation. To give a positive
(non-zero, in particular) mass to the graviton, makes the vector modes
propagate (i.e.\ $c_{V}^{2}>0$). This behavior is quite different
from GR, and, as such, it may lead to some interesting
constraints/phenomenology.

\begin{acknowledgments}
We thank Kurt Hinterbichler, Chunshan Lin and Mark Trodden for useful comments.
The work of A.E.G. and S.M. was supported by WPI Initiative, MEXT, Japan. S.M. also acknowledges the support by Grant-in-Aid for Scientific Research 24540256 and 21111006. 
\end{acknowledgments}


\begin{thebibliography}{99}

\bibitem{Fierz:1939ix} 
  M.~Fierz and W.~Pauli,
  Proc.\ Roy.\ Soc.\ Lond.\ A {\bf 173}, 211 (1939).

\bibitem{Boulware:1973my} 
  D.~G.~Boulware and S.~Deser,
  Phys.\ Rev.\ D {\bf 6}, 3368 (1972).

\bibitem{deRham:2010ik}
  C.~de Rham, G.~Gabadadze, 
  Phys.\ Rev.\  {\bf D82}, 044020 (2010).
  [arXiv:1007.0443 [hep-th]].

\bibitem{deRham:2010kj} 
  C.~de Rham, G.~Gabadadze and A.~J.~Tolley,
  Phys.\ Rev.\ Lett.\  {\bf 106}, 231101 (2011)
  [arXiv:1011.1232 [hep-th]].

\bibitem{Gumrukcuoglu:2011ew} 
  A.~E.~Gumrukcuoglu, C.~Lin and S.~Mukohyama,
  JCAP {\bf 1111}, 030 (2011)
  [arXiv:1109.3845 [hep-th]].

\bibitem{DeFelice:2012mx} 
  A.~De Felice, A.~E.~Gumrukcuoglu and S.~Mukohyama,
  Phys.\ Rev.\ Lett.\  {\bf 109}, 171101 (2012)
  [arXiv:1206.2080 [hep-th]].

\bibitem{Fasiello:2012rw} 
  M.~Fasiello and A.~J.~Tolley,
  JCAP {\bf 1211}, 035 (2012)
  [arXiv:1206.3852 [hep-th]].

\bibitem{Higuchi:1986py} 
  A.~Higuchi,
  Nucl.\ Phys.\ B {\bf 282}, 397 (1987).

\bibitem{Chamseddine:2011bu} 
  A.~H.~Chamseddine and M.~S.~Volkov,
  Phys.\ Lett.\ B {\bf 704}, 652 (2011)
  [arXiv:1107.5504 [hep-th]].
 
\bibitem{D'Amico:2011jj} 
  G.~D'Amico, C.~de Rham, S.~Dubovsky, G.~Gabadadze, D.~Pirtskhalava and A.~J.~Tolley,
  Phys.\ Rev.\ D {\bf 84}, 124046 (2011)
  [arXiv:1108.5231 [hep-th]].

\bibitem{Koyama:2011xz} 
  K.~Koyama, G.~Niz and G.~Tasinato,
  Phys.\ Rev.\ Lett.\  {\bf 107}, 131101 (2011)
  [arXiv:1103.4708 [hep-th]].

\bibitem{Koyama:2011yg} 
  K.~Koyama, G.~Niz and G.~Tasinato,
  Phys.\ Rev.\ D {\bf 84}, 064033 (2011)
  [arXiv:1104.2143 [hep-th]].

\bibitem{Gratia:2012wt} 
  P.~Gratia, W.~Hu and M.~Wyman,
  Phys.\ Rev.\ D {\bf 86}, 061504 (2012)
  [arXiv:1205.4241 [hep-th]];
  M.~Wyman, W.~Hu and P.~Gratia,
  Phys.\ Rev.\ D {\bf 87}, 084046 (2013)
  [arXiv:1211.4576 [hep-th]].

\bibitem{Kobayashi:2012fz} 
  T.~Kobayashi, M.~Siino, M.~Yamaguchi and D.~Yoshida,
  Phys.\ Rev.\ D {\bf 86}, 061505 (2012)
  [arXiv:1205.4938 [hep-th]].

\bibitem{Volkov:2012cf} 
  M.~S.~Volkov,
  Phys.\ Rev.\ D {\bf 86}, 061502 (2012)
  [arXiv:1205.5713 [hep-th]].
 
\bibitem{Gumrukcuoglu:2012aa} 
  A.~E.~Gumrukcuoglu, C.~Lin and S.~Mukohyama,
  Phys.\ Lett.\ B {\bf 717}, 295 (2012)
  [arXiv:1206.2723 [hep-th]].

\bibitem{DeFelice:2013awa} 
  A.~De Felice, A.~E.~Gumrukcuoglu, C.~Lin and S.~Mukohyama,
  JCAP {\bf 1305}, 035 (2013)
  [arXiv:1303.4154 [hep-th]].

\bibitem{LorentzViolation} 
  V.~A.~Rubakov,
  hep-th/0407104;
%
  S.~L.~Dubovsky,
  JHEP {\bf 0410}, 076 (2004)
  [hep-th/0409124];
%
  V.~A.~Rubakov and P.~G.~Tinyakov,
  Phys.\ Usp.\  {\bf 51}, 759 (2008)
  [arXiv:0802.4379 [hep-th]];
%
  D.~Comelli, M.~Crisostomi, F.~Nesti and L.~Pilo,
  Phys.\ Rev.\ D {\bf 86}, 101502 (2012)
  [arXiv:1204.1027 [hep-th]];
%
  D.~Comelli, F.~Nesti and L.~Pilo,
  arXiv:1302.4447 [hep-th];
%
  D.~Comelli, F.~Nesti and L.~Pilo,
  arXiv:1305.0236 [hep-th];
%
  C.~Lin,
  arXiv:1305.2069 [hep-th];
%
  C.~Lin,
  arXiv:1307.2574 [hep-th];
%
  D.~Comelli, F.~Nesti and L.~Pilo,
  arXiv:1307.8329 [hep-th].

\bibitem{Hassan:2011zd} 
  S.~F.~Hassan and R.~A.~Rosen,
  JHEP {\bf 1202}, 126 (2012)
  [arXiv:1109.3515 [hep-th]].

\bibitem{Hassan:2011ea} 
  S.~F.~Hassan and R.~A.~Rosen,
  JHEP {\bf 1204}, 123 (2012)
  [arXiv:1111.2070 [hep-th]].

\bibitem{Hinterbichler:2012cn} 
  K.~Hinterbichler and R.~A.~Rosen,
  JHEP {\bf 1207}, 047 (2012)
  [arXiv:1203.5783 [hep-th]].

\bibitem{vonStrauss:2011mq} 
  M.~von Strauss, A.~Schmidt-May, J.~Enander, E.~Mortsell and S.~F.~Hassan,
  JCAP {\bf 1203}, 042 (2012)
  [arXiv:1111.1655 [gr-qc]].

\bibitem{Comelli:2011zm} 
  D.~Comelli, M.~Crisostomi, F.~Nesti and L.~Pilo,
  JHEP {\bf 1203}, 067 (2012)
  [Erratum-ibid.\  {\bf 1206}, 020 (2012)]
  [arXiv:1111.1983 [hep-th]].

\bibitem{Akrami:2012vf} 
  Y.~Akrami, T.~S.~Koivisto and M.~Sandstad,
  JHEP {\bf 1303}, 099 (2013)
  [arXiv:1209.0457 [astro-ph.CO]].

\bibitem{Comelli:2012db} 
  D.~Comelli, M.~Crisostomi and L.~Pilo,
  JHEP {\bf 1206}, 085 (2012)
  [arXiv:1202.1986 [hep-th]].

\bibitem{DeFelice:2013nba} 
  A.~De Felice, T.~Nakamura and T.~Tanaka,
  arXiv:1304.3920 [gr-qc].

\bibitem{Huang:2012pe} 
  Q.~-G.~Huang, Y.~-S.~Piao and S.~-Y.~Zhou,
  Phys.\ Rev.\ D {\bf 86}, 124014 (2012)
  [arXiv:1206.5678 [hep-th]].

\bibitem{Wu:2013ii} 
  D.~-J.~Wu, Y.~-S.~Piao and Y.~-F.~Cai,
  Phys.\ Lett.\ B {\bf 721}, 7 (2013)
  [arXiv:1301.4326 [hep-th]].

\bibitem{Hinterbichler:2013dv} 
  K.~Hinterbichler, J.~Stokes and M.~Trodden,
  Phys.\  Lett.\ B {\bf 725}, , 1 (2013)
  [arXiv:1301.4993 [astro-ph.CO]].

\bibitem{Leon:2013qh} 
  G.~Leon, J.~Saavedra and E.~N.~Saridakis,
  Class.\ Quant.\ Grav.\  {\bf 30}, 135001 (2013)
  [arXiv:1301.7419 [astro-ph.CO]].

\bibitem{Gumrukcuoglu:2013nza} 
  A.~E.~Gumrukcuoglu, K.~Hinterbichler, C.~Lin, S.~Mukohyama and M.~Trodden,
  Phys.\ Rev.\ D {\bf 88}, 024023 (2013)
  [arXiv:1304.0449 [hep-th]].

\bibitem{DeFelice:2013tsa} 
  A.~De Felice and S.~Mukohyama,
  arXiv:1306.5502 [hep-th].

\bibitem{D'Amico:2012zv} 
  G.~D'Amico, G.~Gabadadze, L.~Hui and D.~Pirtskhalava,
  Phys.\ Rev.\ D {\bf 87}, 064037 (2013)
  [arXiv:1206.4253 [hep-th]].

\bibitem{D'Amico:2013kya} 
  G.~D'Amico, G.~Gabadadze, L.~Hui and D.~Pirtskhalava,
  arXiv:1304.0723 [hep-th].

\bibitem{Mukohyama:2013raa} 
  S.~Mukohyama,
  arXiv:1309.2146 [hep-th].

\bibitem{Horndeski:1974wa} 
  G.~W.~Horndeski,
  Int.\ J.\ Theor.\ Phys.\  {\bf 10}, 363 (1974).

\bibitem{Deffayet:2009mn} 
  C.~Deffayet, S.~Deser and G.~Esposito-Farese,
  Phys.\ Rev.\ D {\bf 80}, 064015 (2009)
  [arXiv:0906.1967 [gr-qc]].

\bibitem{Charmousis:2011bf} 
  C.~Charmousis, E.~J.~Copeland, A.~Padilla and P.~M.~Saffin,
  Phys.\ Rev.\ Lett.\  {\bf 108}, 051101 (2012)
  [arXiv:1106.2000 [hep-th]].

\bibitem{Gabadadze:2012tr} 
  G.~Gabadadze, K.~Hinterbichler, J.~Khoury, D.~Pirtskhalava and M.~Trodden,
  Phys.\ Rev.\ D {\bf 86}, 124004 (2012)
  [arXiv:1208.5773 [hep-th]].

\end{thebibliography}
\end{document}